%% file: acat-mctuning-jhep.tex
\documentclass{JHEP3}
\usepackage{hepunits,hepnames}
\usepackage{booktabs}
\usepackage{graphicx,subfig}
\usepackage{xspace}
\usepackage{cite}
\usepackage{microtype}
\usepackage{snapshot}
\usepackage{footnote}

\usepackage{caption}
\author{\speaker{Andy Buckley}\\
        Institute for Particle Physics Phenomenology, Durham University, UK\\
        E-mail: \email{andy.buckley@durham.ac.uk}}
\author{Hendrik Hoeth\\
        Department of Theoretical Physics, Lund University, Sweden}
\author{Holger Schulz\\
        Physics Department, Berlin Humboldt University, Germany}
\author{Jan Eike von Seggern\\
        Physics Department, TU-Dresden, Germany}

\title{Monte Carlo event generator validation and\\ tuning for the LHC}

\newcommand{\Rivet}{Rivet\xspace}
\newcommand{\Professor}{Professor\xspace}
\newcommand{\HZTool}{\textsc{HZTool}\xspace}
\newcommand{\HepData}{\textsc{HepData}\xspace}
\newcommand{\Delphi}{\textsc{Delphi}\xspace}
\newcommand{\Aleph}{\textsc{Aleph}\xspace}
\newcommand{\Opal}{\textsc{Opal}\xspace}
\newcommand{\DZero}{D\O\xspace}
\newcommand{\CDF}{\textsc{CDF}\xspace}

\newcommand{\ee}{\ensuremath{\eplus\eminus}\xspace}
\newcommand{\Order}[1]{\ensuremath{\mathcal{O}(#1)}\xspace}

\preprint{IPPP/09/11\\DCPT/09/22\\MCnet/09/03}

\abstract{We summarise the motivation for, and the status of, the tools developed
  by CEDAR/MCnet for validating and tuning Monte Carlo event generators for the
  LHC against data from previous colliders. We then present selected preliminary
  results from studies of event shapes and hadronisation observables from
  $\eplus\eminus$ colliders, and of minimum bias and underlying event observables
  from the Tevatron, and comment on the approach needed with early LHC data to
  best exploit the potential for new physics discoveries at the LHC in the next
  few years.}

\begin{document}

\input{acat-mctuning-body}

\bibliographystyle{h-physrev3}
{\raggedright
  \bibliography{acat}
}

\end{document}

%% file: acat-mctuning-body.tex
\section{Introduction}

The LHC is designed to discover what lies beyond the \TeV{} scale, so tantalisingly
probed by the Tevatron experiments. The most obvious theoretically motivated
candidates for discovery are the Higgs boson and supersymmetry, but unitarity
and renormalisation arguments mean that it is extremely likely that we will find
\emph{something} new.

One thing that we are \emph{absolutely} certain to observe is the Standard
Model! In particular, the LHC will be a probe of QCD machine as it has never
been seen before: the proton will be probed in regions of high momentum transfer
and low Bj\"orken $x$ (requiring a new understanding of parton densities and
hence new PDF fits), jets above \unit{1}{\TeV} will be seen, and the behaviour
of the $pp$ total cross-section and multiple parton interactions will be
measured at values of $\sqrt{s}$ where current data offer little constraint. It
is certain that the SM will need to be measured and understood in this new
regime before any new physics discovery can be claimed with confidence.

Key to the process of developing new physics analyses is the simulation of both
background and signal events. Particularly with the rise of multivariate
methods, such as have been discussed in other ACAT parallel sessions, the
discrimination between signal and background is often tuned to predictions from
Monte Carlo event generators. While published analyses must be virtually
independent of such modelling assumptions, the accuracy of the physics
description provided by the simulation codes is crucial for efficient
exploitation of LHC data. In the first part of this talk, we summarise the
current state of efforts to systematically check the validity of MC generator
simulations, and to improve their performance by systematic parameter tuning. In
the second part, we will focus on the underlying event as an area of physics
whose MC description can be improved before LHC running by use of Tevatron data,
and which must be re-tuned to early LHC data when available, in order to make
the most of LHC BSM studies in the early years of the collider.

\section{Event generator tuning and validation tools}

Despite the importance of MC simulation to the development of LHC physics
studies, there has until recently been a dearth of coordinated MC validation
studies. That is, while plenty of individual LHC physics analyses have
considered private plots of a generator's predictions, there has not been a
study broad enough to provide side-by-side displays of how different simulation
codes perform, both with respect to each other and to data from previous
experiments. This may be due to the awkwardness of the task: all generators are
run in a different way, with different steering parameters; and ensuring that
the event records produced by them can be manipulated to provide data which may
be compared to that from existing experiments is a task ill-suited to
experimental physicists under pressure to produce plots for one specific
process. However, the task is important, as changes in generator parameter
choices can profoundly affect their predictions, and a tuning which appears good
for one observable may be unphysically awful for another.

The CEDAR project, on which we reported at the previous
ACAT\,\cite{Buckley:2007hi}, was established to provide manpower to address this
issue by developing tools for MC validation. Since then, CEDAR has been
integrated into the MCnet EU research network, which is ideal for sharing
developments between generator developers and LHC experimentalists. The main
tools developed by (and now being used by) CEDAR are the validation tool
\emph{\Rivet{}}, and the \emph{\Professor{}} tuning system. We have reported on these
systems before, so our description will accordingly be brief and focussed on
recent developments: anyone to whom these tools are entirely new is advised to
check refs.~\cite{Waugh:2006ip,Buckley:2007hi,Buckley:2008vh}.

\subsection{Rivet}

\Rivet is an analysis framework for MC generator validation, intended originally
to be a modern, C++\footnote{Not entirely an oxymoron\dots but certainly not a
  tautology!}  successor to the venerable \HZTool system. Key design features of
\Rivet are that the HepMC event record is the only data source (hence providing
isolation from the temptation to query generator internals), that CPU-intensive
computations of observables from stable particles are automatically cached for
each event and shared between analyses, and that the reference data may be
automatically exported to flat data files from the \HepData archive\footnote{An
  earlier CEDAR project, described in the previous ACAT proceedings, involved
  the upgrade of \HepData's database system and addition of a Java data model to
  make such exporting possible.}\,\cite{Buckley:2006np}. The reference data is
also used to define the binnings of MC histograms, automatically ensuring that
there is no problem with synchronising arrays of bin edge positions.

Internally, \Rivet analyses can be programmed using a very clean C++ API which
isolate physicist users from the details of object memory and life-cycle
management. External analyses may be built as shared libraries and loaded at
runtime by a simple ``plugin'' system. A Python interface to the \Rivet and
HepMC libraries, implemented using the SWIG wrapper generator, is used to
provide a very user-friendly command-line interface to \Rivet analyses,
including analysis metadata querying. At present, there are roughly 40 key
analyses from LEP, SLD and Tevatron experiments in Rivet: most development
effort is focussed on adding QCD analyses from ``missing'' colliders such as
ISR, SPPS, CLEO, and the \Pbottom-factories. HERA analyses will primarily remain in
\HZTool, updated with a HepMC input layer and histogram output compatible with
\Rivet.

The current stable version of \Rivet is 1.1.2 (with a 1.1.3 patch release
expected soon). The 1.2.0 version will provide much-improved histogramming,
after which the analysis infrastructure will be essentially complete and all
effort will be on adding more analyses and exploiting \Rivet for more advanced
generator validation studies.

\subsection{Professor}

\Professor is an extension of the \textsc{Delphi} generator tuning system,
developed by a collaboration between MCnet, TU-Dresden, and Berlin Humboldt
University. Unlike either the intrinsically sub-optimal ``by-eye'' tunings
commonly delegated to unfortunate graduate students, or brute-force tunings ---
which rapidly fail to scale to large parameter spaces, even in these days of
grid computing --- it is based on parameterising the response of MC observable
bins to correlated shifts in generator parameters via a polynomial, usually
second order. Accordingly, there is an assumption that the generator responds in
a sufficiently smooth way to parameter variations, but in practice this proves
to be true --- at least when the bin variations are combined together to compute
some goodness of fit function (GoF), e.g. a heuristic $\chi^2$, against
reference data. The parameterisation is determined by randomly sampling
parameter vectors from the parameter hypercube and running the generator at each
sampled point via a batch cluster or the LCG grid: a singular value
decomposition is used to deterministically implement the ``pseudoinverse'' which
determines each bin's best polynomial coefficients according to a least squares
definition. It is usually possible to factorise the 30 or so main interesting
parameters of a generator like Pythia into semi-independent groups of 5--10, and
the scaling of the minimum number of runs --- generally dependent on the
polynomial order --- gives $N_\text{min} \sim \Order{100}$. To obtain some
estimate of the systematic error introduced by the procedure, we actually sample
$N \sim 3N_\text{min}$ runs and then construct $\Order{100}$ different sets of
parameterisations by randomly choosing $N_\text{fit}$ of those runs, where
$N_\text{min} < N_\text{fit} < N$. Finally, the Minuit numerical minimiser is
used to minimise the parameterised GoFs and produce a set of (hopefully
consistent) predicted optimal parameter sets. This procedure has the desirable
feature of combined performance and tractability: as $\Order{\text{10M}}$ events
may be needed for each parameter point (typically $\approx$ 2--3 CPU days),
minimising an analytic function created from hundreds of such runs in parallel
is \emph{vastly} preferable to running thousands of serial Minuit runs.

Several things about this procedure are worth noting: first, the choice of
parameter ranges is the responsibility of the user and must be based on an
understanding of the generator. Too wide choices will be insufficiently
sensitive to the interesting region; too narrow and the predicted minimum may be
outside the sampled ranges\footnote{At least in this case, the course is clear:
  extend the range and run the generator for another weekend.}. Second, the
choice of GoF function is flexible, most significantly in that typically one
will consider certain distributions to be more significant than others and give
them an extra \emph{weight} accordingly. The choice of numeric weights to
maintain a balanced minimisation is something of an art form, illustrating the
ever-useful rule that when one demands optimal solutions, he should be careful
about exactly what he asks for.

An extra technical point, which usually leads to questions from MC authors
accustomed to integrating awkward functions, is the random sampling of the
parameter hypercube. So far we have always sampled uniformly on the parameter
space, without evident bias. For particularly non-linear parameters, non-linear
sampling or parameter transformations could be used: in essence we want to
sample according to the ``prior reasonableness'' of the parameters, and in the
absence of other information a flat prior is the natural choice (and not a
dangerous one since the ranges are bounded). However, this is another area where
knowledge of the generator is useful: tunings are most definitely best done in
collaboration with the generator authors.

Finally, note that there is nothing MC-specific about this method: it is a
general method for minimising very expensive functions where there is no \emph{a
  priori} estimate for the functional form. Accordingly, within HEP the
\Professor approach has been adapted to fitting the top quark mass and to
choosing the parameters of unintegrated PDFs for the CCFM
shower formalism\,\cite{Knutsson:2008qs}.

\section{Validation and tuning of MC simulations to LEP and Tevatron data}

Having summarised the Professor method for Monte Carlo tunings, we will now
discuss the tuning of Pythia~6 parameters to $\eplus\eminus$ and Tevatron data,
constraining the parameters of the initial and final state parton showers,
hadronisation, and multiple parton interactions (MPI) model. We will only
consider the tune to Pythia's traditional virtuality-ordered parton shower and
``old'' MPI model --- the full tune details of the newer $p_\perp$-ordered
shower and more complex interleaved ISR/MPI model will be discussed in a
forthcoming publication.

The number of major steering parameters in Pythia~6 is approximately 30, with
the majority being associated with aspects of hadronisation. While practically
achievable, albeit not easily, we would rather not have to tune in 30 dimensions
since the likelihood of undersampling or of the minimiser failing to find the
global minimum is relatively high. Fortunately, the parameters can be roughly
factorised into sets of less than 10 which can be tuned almost independently:
for example, the flavour composition of the final state particles is irrelevant
when calculating event shape observables and hence the kinematic/flavour
parameters can be treated independently. Similarly the treatment of tensor
mesons and relative production of different diquark spin states.

\subsection{Tuning of FSR and hadronisation to $\eplus\eminus$ data}

\begin{table}[t]
  \centering


  \begin{tabular}{lr@{.}lr@{.}ll}
    \toprule
    & \multicolumn{2}{c}{default}
    & \multicolumn{2}{c}{new tune} & \\
    \midrule
    \multicolumn{5}{l}{\textbf{Kinematic parameters}} & \\
    MSTJ(11) & \multicolumn{2}{l}{4} & \multicolumn{2}{l}{5} & Frag fn.\\
    PARJ(21) & 0&36  & 0&325   & $\sigma_q$ \\
    PARJ(41) & 0&3   & 0&5     & Lund $a$        \\
    PARJ(42) & 0&58  & 0&6     & Lund $b$        \\
    PARJ(47) & 1&0   & 0&67    & $r_b$      \\
    PARJ(81) & 0&29  & 0&29    & $\Lambda_\text{QCD}$  \\
    PARJ(82) & 1&0   & 1&65    & PS cut-off \\
    \multicolumn{5}{l}{\textbf{Flavour parameters}} & \\
    PARJ(1)  & 0&1   & 0&073   & di-quark suppression \\
    PARJ(2)  & 0&3   & 0&2     & strange suppression \\
    PARJ(3)  & 0&4   & 0&94    & strange di-quark suppression \\
    PARJ(4)  & 0&05  & 0&032   & spin-1 di-quark suppression \\
    PARJ(11) & 0&5   & 0&31    & spin-1 light meson \\
    PARJ(12) & 0&6   & 0&4     & spin-1 strange meson \\
    PARJ(13) & 0&75  & 0&54    & spin-1 heavy meson \\
    PARJ(25) & 1&0   & 0&63    & $\eta$ suppression \\
    PARJ(26) & 0&4   & 0&12    & $\eta'$ suppression \\
    \bottomrule
  \end{tabular}

  \caption{Parameters used in the tune of Pythia~6 to LEP/SLD data, and their resulting values.}
  \label{tab:eeparams}
\end{table}

\begin{figure}[t]
  \centering
  \subfloat[][]{\includegraphics[width = 0.25\textwidth, trim = 0 -50 0 0]{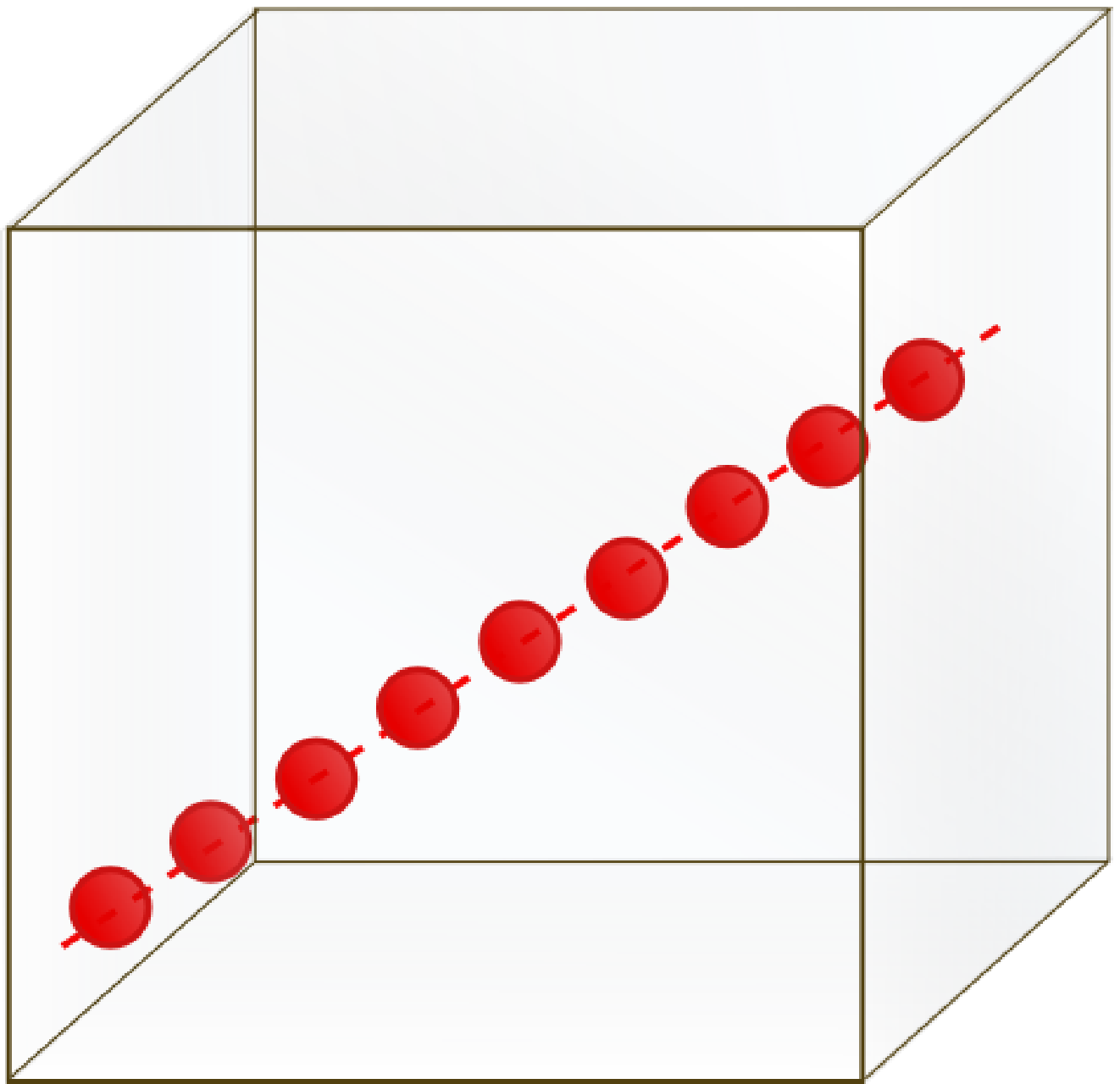}}
  \qquad
  \subfloat[][]{\includegraphics[width = 0.5\textwidth]{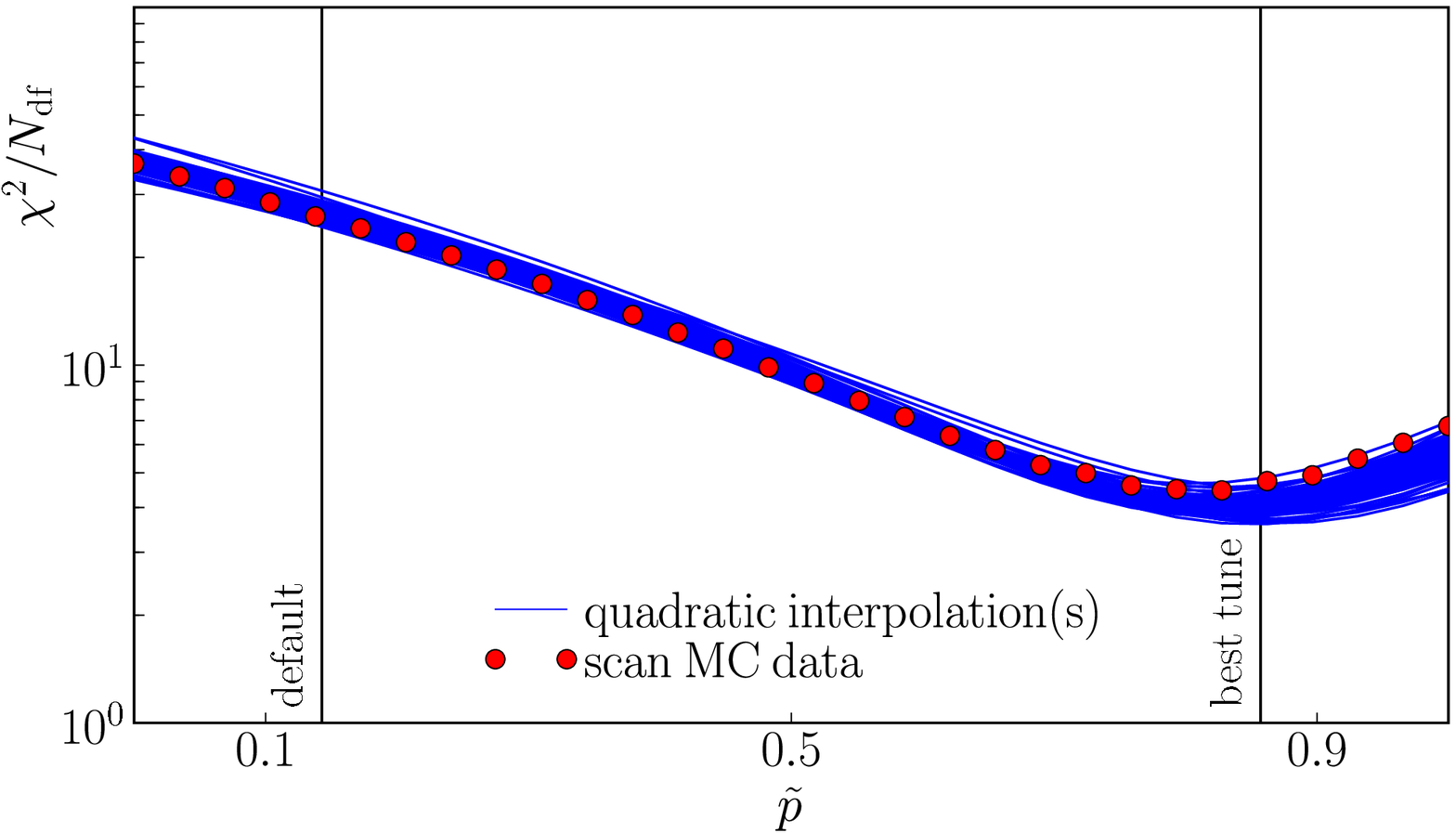}}
  \caption{Pythia 6 ($Q^2$ shower) $\chi^2/N_\text{df}$ variation along a line
    in the parameter hyperspace, as illustrated in (a). The line shown in (b)
    runs between the default and Professor tunes for the flavour parameter
    tuning. The red dots are the true generator $\chi^2$ values, and the blue
    lines an ensemble of parameterisations from the Professor procedure. The
    \Professor result is clearly superior, although it does not match the true
    optimum exactly.}%
  \label{fig:pyflavscan}%
\end{figure}

Observing this approximate independence of certain parameter groups, we begin
our tuning of Pythia~6 with tuning of final state radiation (FSR) and
hadronisation effects at the LEP and SLD \ee colliders. The relevant
parameters are shown in Table~\ref{tab:eeparams}, factorised into ``kinematic''
and ``flavour'' sets. The key sets of distributions to fit are the total charged
multiplicity at LEP by the \Opal collaboration\,\cite{Ackerstaff:1998hz}, 
identified particle multiplicities from the Particle Data
Group\,\cite{Amsler:2008zz}, event shape variables from
\Aleph{}\,\cite{Barate:1996fi} and \Delphi{}\,\cite{Abreu:1996na}, and
\Pbottom-fragmentation from \Delphi{}\,\cite{delphi-2002}. All are influenced by
shower and hadronisation kinematics, but only the second is sensitive to the
flavour parameters. However, if we only consider \emph{ratios} of identified
particle rates, they are roughly independent of kinematic parameters.  Hence our
\ee tuning is implemented in 2 stages, with the second stage fixing the
parameters determined in the first:
%
\begin{enumerate}
\item Tune flavour parameters to the identified particle rates, relative to the \Ppipm rate;
\item Tune kinematic parameters to event shapes and total charged multiplicity.
\end{enumerate}
%
Each stage of the tune was performed using 300 random runs of 200k events in
$\eplus\eminus$ configuration at the LEP 1 CoM energy of \unit{91.2}{\GeV}.  The
resulting tune, which has been checked for robustness against many choices of
runs and observable weights, are listed in Table~\ref{tab:eeparams}. The mean
charged multiplicity, being a single number of disproportionate importance, was
given an effective weight of 220 (spread between several measurements), while
most other distributions were given weights less than 10 --- most were set to 1.
This boosted weight helps to compensate for the down-weighting that is implicit
in any distributions with a fewer than typical number of bins, and which is
hence most significant for single-bin observables. The robustness of the
approach is illustrated in Figure~\ref{fig:pyflavscan}, which illustrates the
GoF measure as predicted by Professor and as realised by the generator along a
line scan in flavour parameter space between the default and Professor tunes:
the Professor result is clearly near-optimal, while the default appears
relatively arbitrary.

\subsection{Tuning of ISR and MPI to $\Pproton\APproton$ data}

The final state parameters derived in the tunes to $\eplus\eminus$ data can now
be used as a base around which to tune the parameters controlling initial state
effects in hadron collisions. For this we use Tevatron data, primarily from the
\CDF experiment: the \CDF measurement of the \PZ $p_\perp$
spectrum\,\cite{Affolder:1999jh}, the \DZero measurement of dijet azimuthal
angle decorrelation\,\cite{Abazov:2004hm}, and \CDF measurements of the
``underlying event'' (MPI) from both Run~I and
Run~II\,\cite{Affolder:2001xt,Acosta:2001rm,cdf-note9337,cdf-note9351,cdf-leadingjet}. In
general, such tunes are sensitive to the choice of parton density set used: in
this tune we use the Pythia~6 default, the leading order CTEQ5L
fit\,\cite{Lai:1999wy}. We are currently repeating this tuning with PDF sets
useful for LHC experiment production simulations in 2009, and for the new Monte
Carlo specific ``modified LO'' sets such as the LO* and LO** variations on the
MSTW LO PDF sets\,\cite{Sherstnev:2008dm}.

\begin{table}[t]
  \centering
  \begin{tabular}{lr@{.}lr@{.}lr@{.}ll}
    \toprule
    & \multicolumn{2}{c}{default}
    & \multicolumn{2}{c}{tune DW}
    & \multicolumn{2}{c}{new tune} & \\
    \midrule
    PARP(62) & 1&0   &  1&25  & 2&97  & ISR cut-off \\
    PARP(64) & 1&0   &  0&2   & 0&12  & ISR scale factor for $\alpha_s$ \\
    PARP(67) & 4&0   &  2&5   & 2&74  & max. virtuality \\
    PARP(82) & 2&0   &  1&9   & 2&1   & $p_{\perp}^0$ \\
    PARP(83) & 0&5   &  0&5   & 0&84  & matter distribution \\
    PARP(84) & 0&4   &  0&4   & 0&5   & matter distribution \\
    PARP(85) & 0&9   &  1&0   & 0&82  & colour connection \\
    PARP(86) & 0&95  &  1&0   & 0&91  & colour connection \\
    PARP(90) & 0&16  &  0&25  & 0&17  & $p_\perp^0$ energy evolution \\
    PARP(91) & 2&0   &  2&1   & 2&0   & intrinsic $k_\perp$ \\
    PARP(93) & 5&0   & 15&0   & 5&0   & intrinsic $k_\perp$ cut-off \\
    \bottomrule
  \end{tabular}
  \caption{Parameters used in the tune of Pythia~6 initial state physics
    to Tevatron data, and their resulting values.}
  \label{tab:ppbarparams}
\end{table}

This time, no factorisation of the parameters listed in
Table~\ref{tab:ppbarparams} is required. The \PZ $p_\perp$ spectrum is sensitive
particularly to the ISR and intrinsic $p_\perp$ parameters, since these are the
only way that transverse momentum can be given to the \PZ boson. If this is
ignored, the fit to MPI observables tends to destroy the \PZ $p_\perp$
description: this was the motivation for the evolution of the Tevatron ``Tune
A'' to ``Tune AW''. Similarly, the ISR/intrinsic $p_\perp$ contributions must be
constrained to the \DZero dijet angular decorrelation --- a measure of how the
initial state effects disturb the back-to-back picture of dijet events --- which
was the motivation for the Tevatron ``Tune DW''. We incorporate these into our
Tevatron tune with weights of 40 and 2 respectively: the latter weight is not
set particularly high because no setting of Pythia~6 seems to describe this data
\emph{very} well, hence it acts as a veto on bad tunes rather than a driver of
very good ones.

The run conditions for this data are more complex than for the \ee case, since
several energies and process types (QCD and Drell-Yan) are involved, and because
to efficiently fill profile histograms binned in leading jet or \PZ $p_\perp$
requires several runs with kinematic ME cuts to be combined. The required
statistics are also larger, varying between 1--2M events per run, as compared to
$\Order{\text{100k}}$ events for the \ee tune. The resulting parameter set,
again checked for robustness, is listed in Table~\ref{tab:ppbarparams}.

It should be noted that the constraint of the $\Pproton\APproton$ cross-section
energy evolution, which is a crucial number for LHC physics, is weakly
constrained by this tune, since the only contributing energies are 1800 and
\unit{1960}{\GeV}; this is remedied by the tunes in our forthcoming publication,
which include data from \unit{630}{\GeV}, and we hope also to include data from
\unit{200}{\GeV} $\Pproton\Pproton$ runs at RHIC and earlier hadron colliders.


\section{Comparisons of generators/tunes}

The important thing about a generator tune, of course, is whether or not it
describes the data. This is not necessarily guaranteed, even with a procedure
like Professor: there is a subjectivity in the choice of observables and the
weights they are afforded, and --- more fundamentally --- there is no guarantee
that the generator/model is capable of describing the data well at several
energies or even at one energy. Fortunately, Pythia~6 proves itself up to the challenge in this
case. In this section, we shall show a small selection of the distributions to
which Pythia~6 has been tuned, comparing to other tunes of the same generator,
and finally compare the quality of the data description offered by the Professor
tune of Pythia~6 to that from other shower/hadronisation codes in various states
of tunelessness.

\subsection{Comparisons of Pythia~6 tunes}

\begin{figure}[hb]%
  \centering
  \subfloat[][]{\includegraphics[width=0.45\textwidth]{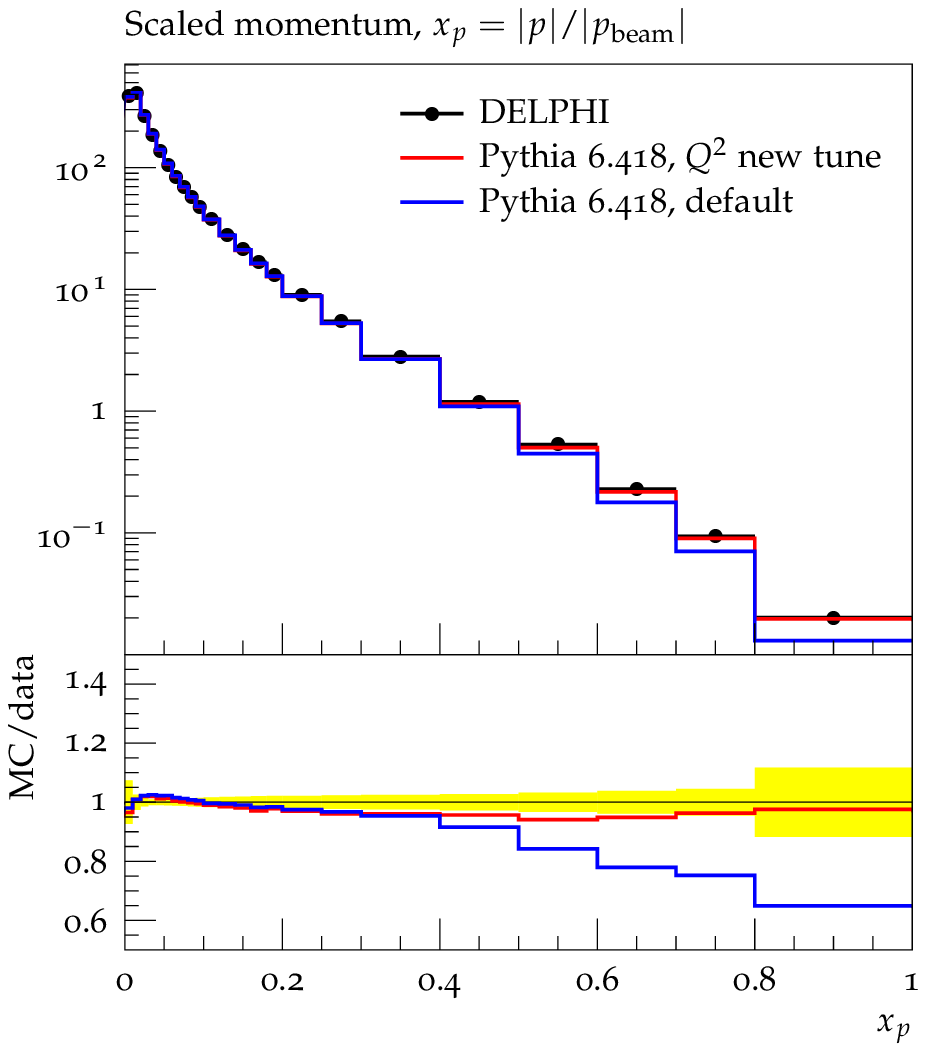}}%
  \qquad
  \subfloat[][]{\includegraphics[width=0.45\textwidth]{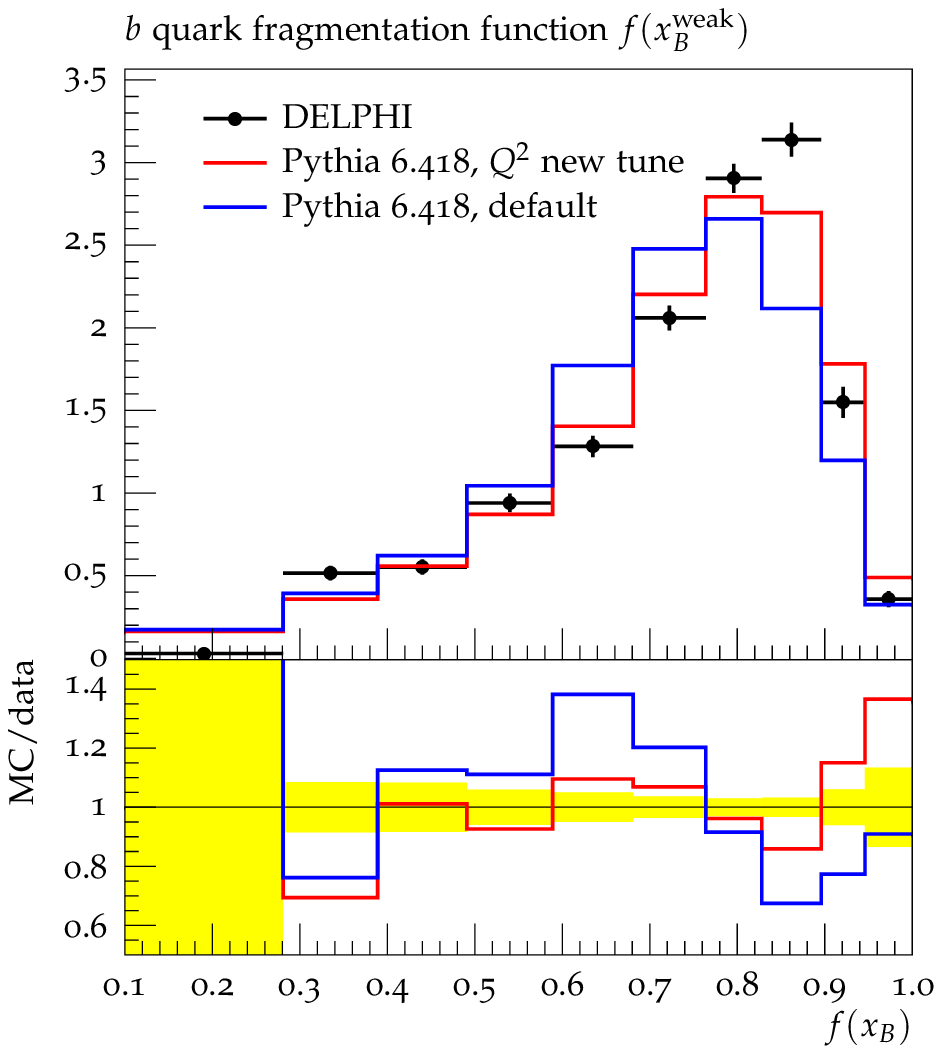}}%
  \caption{Pythia 6 ($Q^2$ shower) comparative performance on LEP scaled
    momentum and \Pbottom-quark fragmentation function data.}%
  \label{fig:pyeeobs}%
\end{figure}

In Figures~\ref{fig:pyeeobs} and \ref{fig:pyppbarobs}, the improvements of the
Professor tune with respect to the default tune are shown for selected
$\eplus\eminus$ and $\Pproton\APproton$ observables. The improvements in the
$\eplus\eminus$ case are small --- unsurprising, since the default Pythia~6 tune
is based on the original \Delphi version of the \Professor procedure --- but
significant, especially in the case of the \Pbottom-quark fragmentation
function. The Tevatron observables see improvements over the ``AW'' tune in
several areas, particularly the minimum bias and Drell-Yan $\langle p_\perp
\rangle$ vs. $N_\text{charged}$ distributions, and are a major improvement over
the default tune (not shown).

The traces shown in the figures also include two tunes of the newer Pythia
$p_\perp$-ordered shower and new MPI model, which are of interest because a)
they demonstrate that the newer MPI model is capable of describing the bump at
$\sim \unit{20}{\GeV}$ in the \CDF 2008 leading jets analysis and b) the Atlas
tune is badly wrong in many areas, especially the description of all Drell-Yan
UE data. The problems of the Atlas tune, and the capabilities of the new MPI
model will be addressed in the forthcoming publication of the \Professor tunes
of both Pythia~6 shower/MPI models.
\vspace*{2cm}

\begin{figure}[p]%
  \centering
  \subfloat[][]{\includegraphics[width=0.45\textwidth]{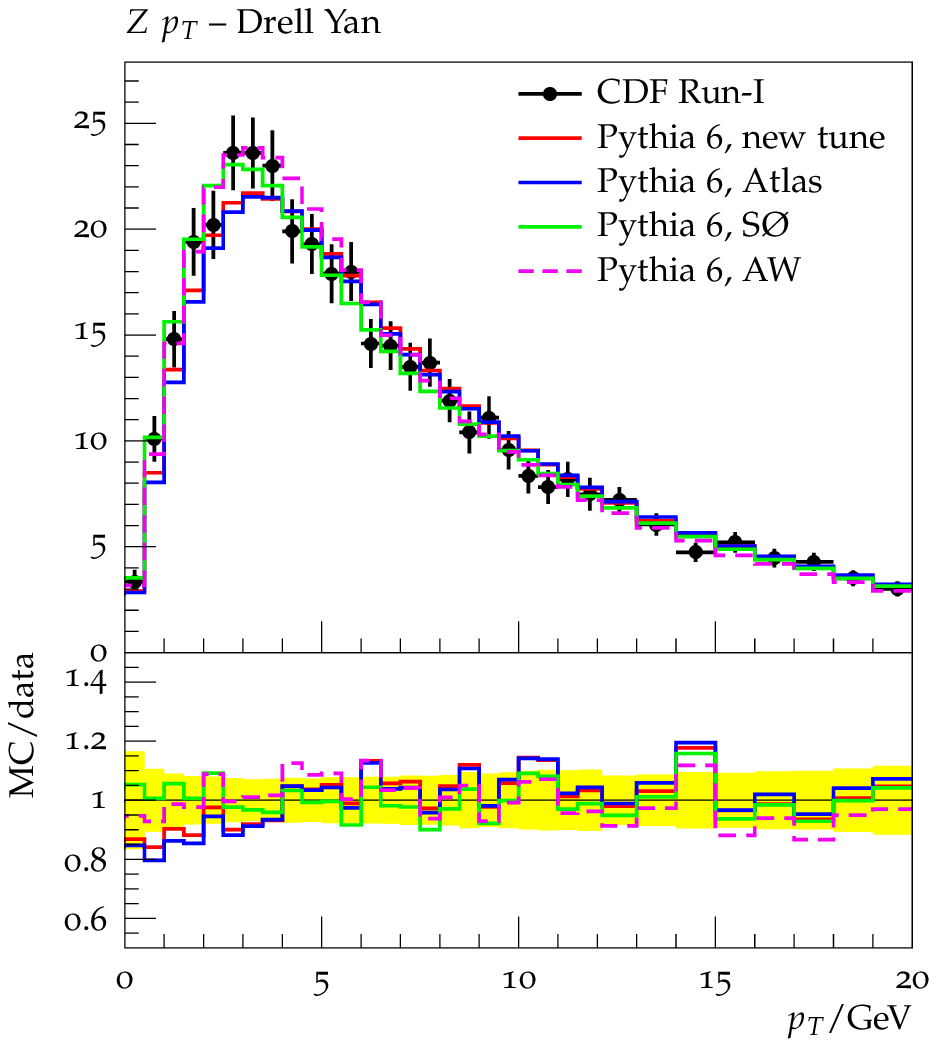}}%
  \qquad
  \subfloat[][]{\includegraphics[width=0.45\textwidth]{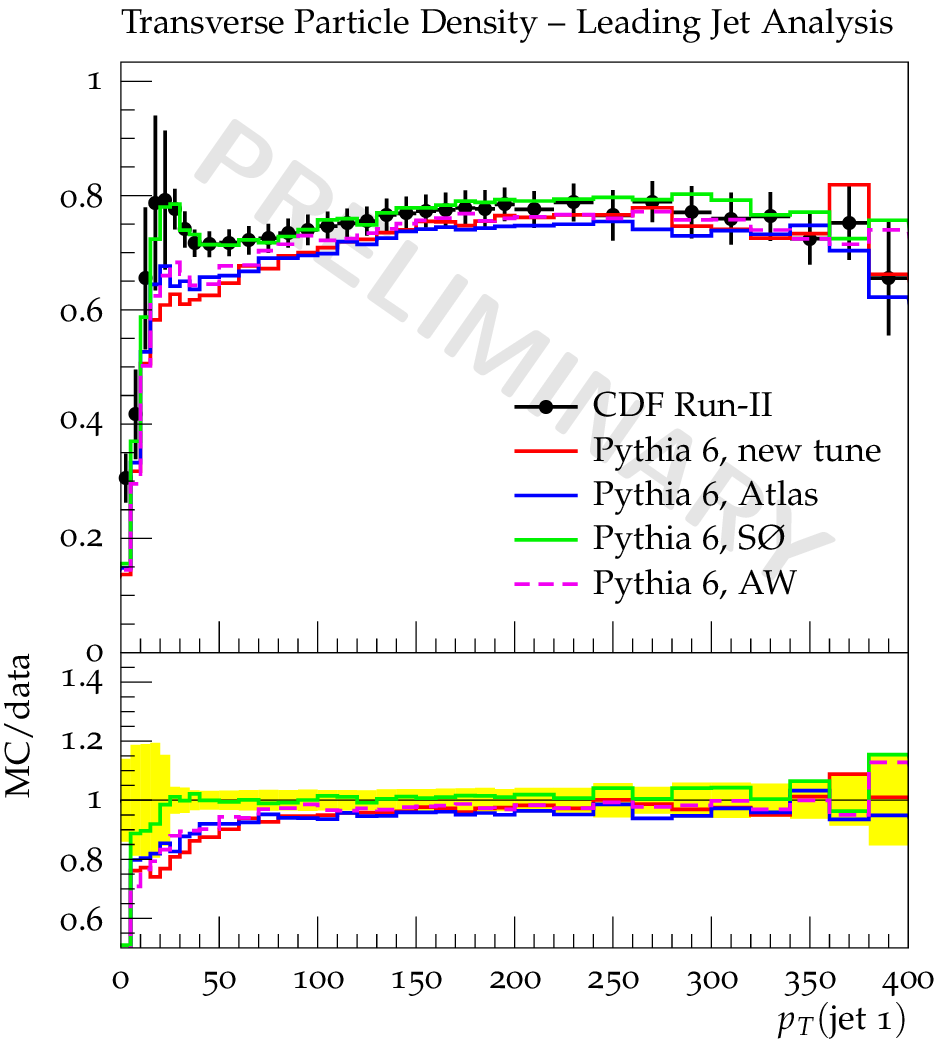}}%

  \subfloat[][]{\includegraphics[width=0.45\textwidth]{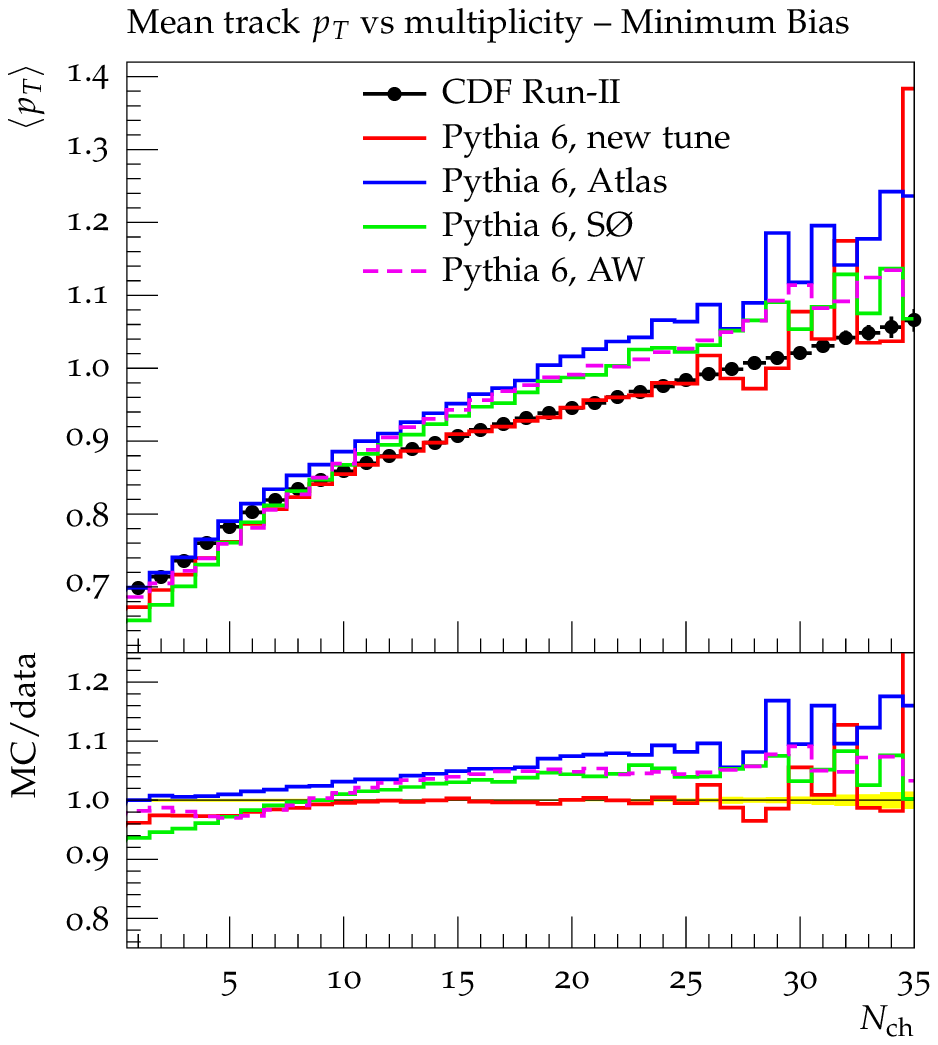}}%
  \qquad
  \subfloat[][]{\includegraphics[width=0.45\textwidth]{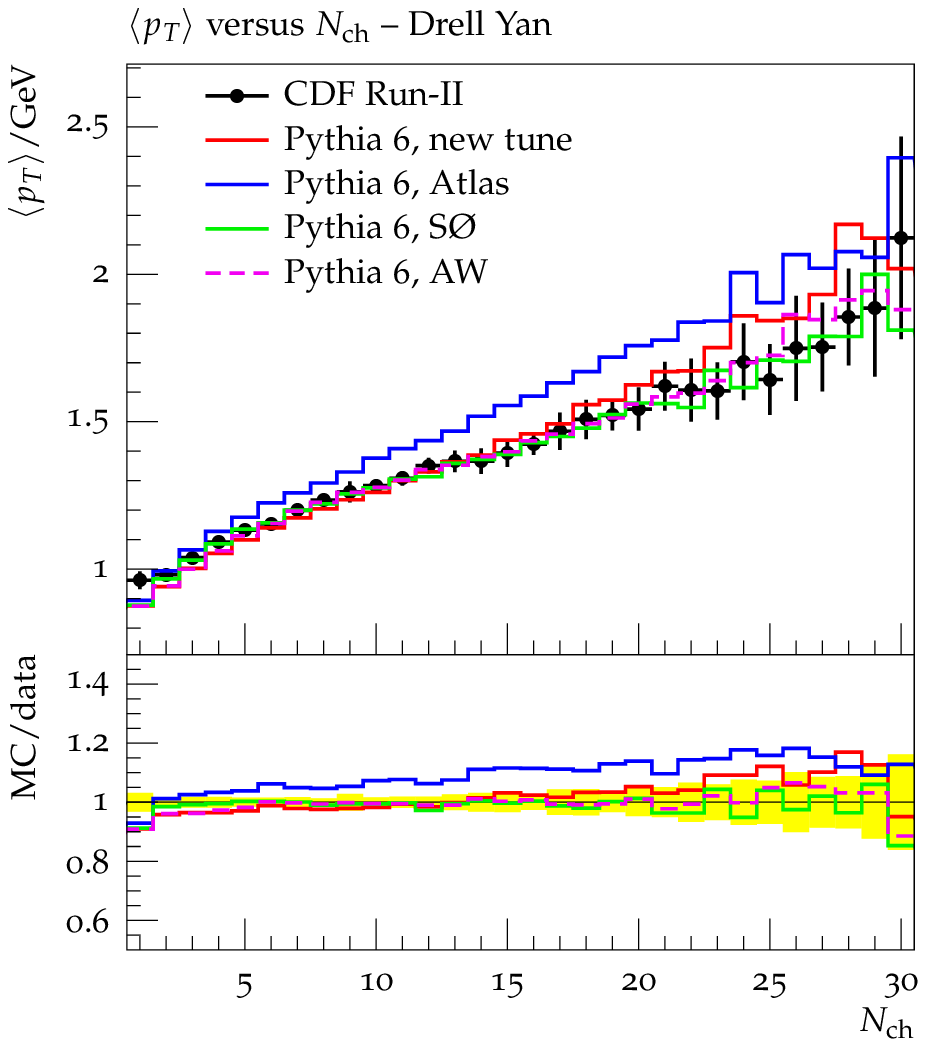}}%
  \caption{Pythia 6 ($Q^2$ shower, ``old'' MPI) comparative performance on Tevatron
    minimum bias and underlying event data at 1800 and \unit{1960}{\GeV}.}%
  \label{fig:pyppbarobs}%
\end{figure}

\begin{figure}[p]%
  \ContinuedFloat
  \centering
  \subfloat[][]{\includegraphics[width=0.45\textwidth]{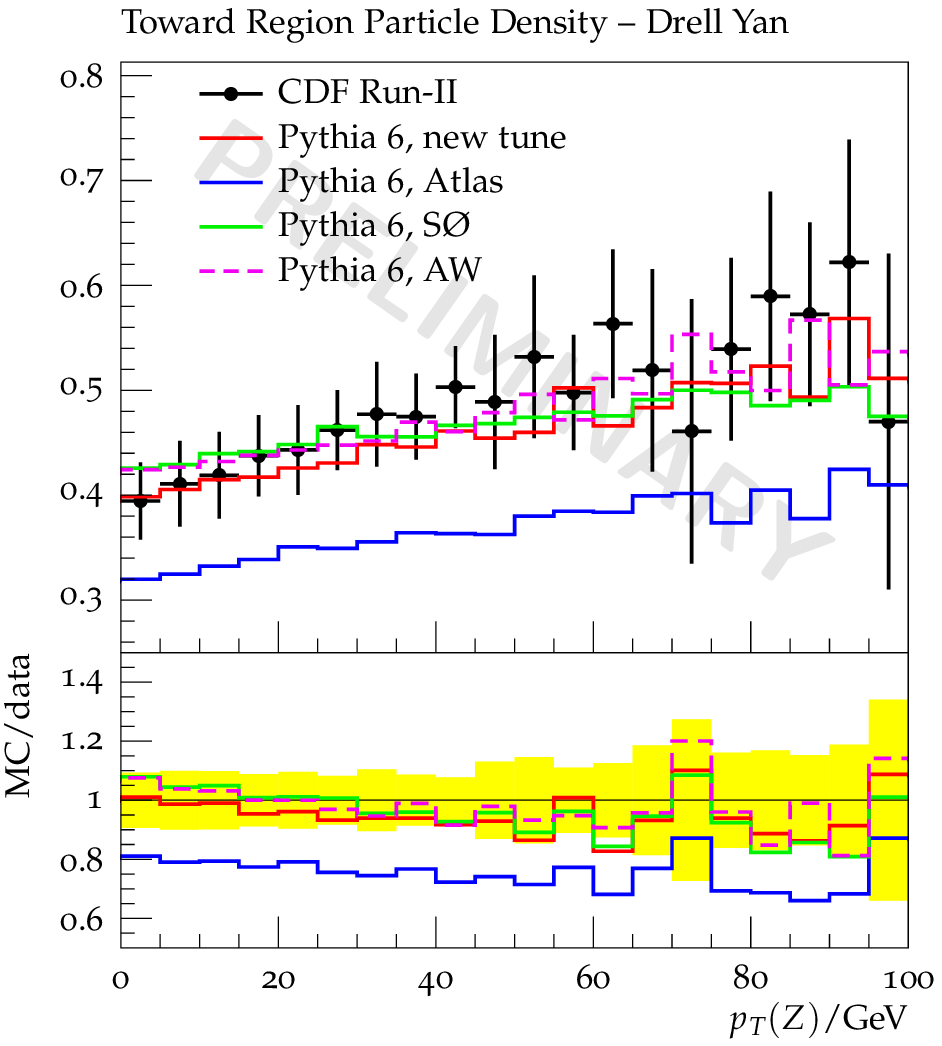}}%
  \qquad
  \subfloat[][]{\includegraphics[width=0.45\textwidth]{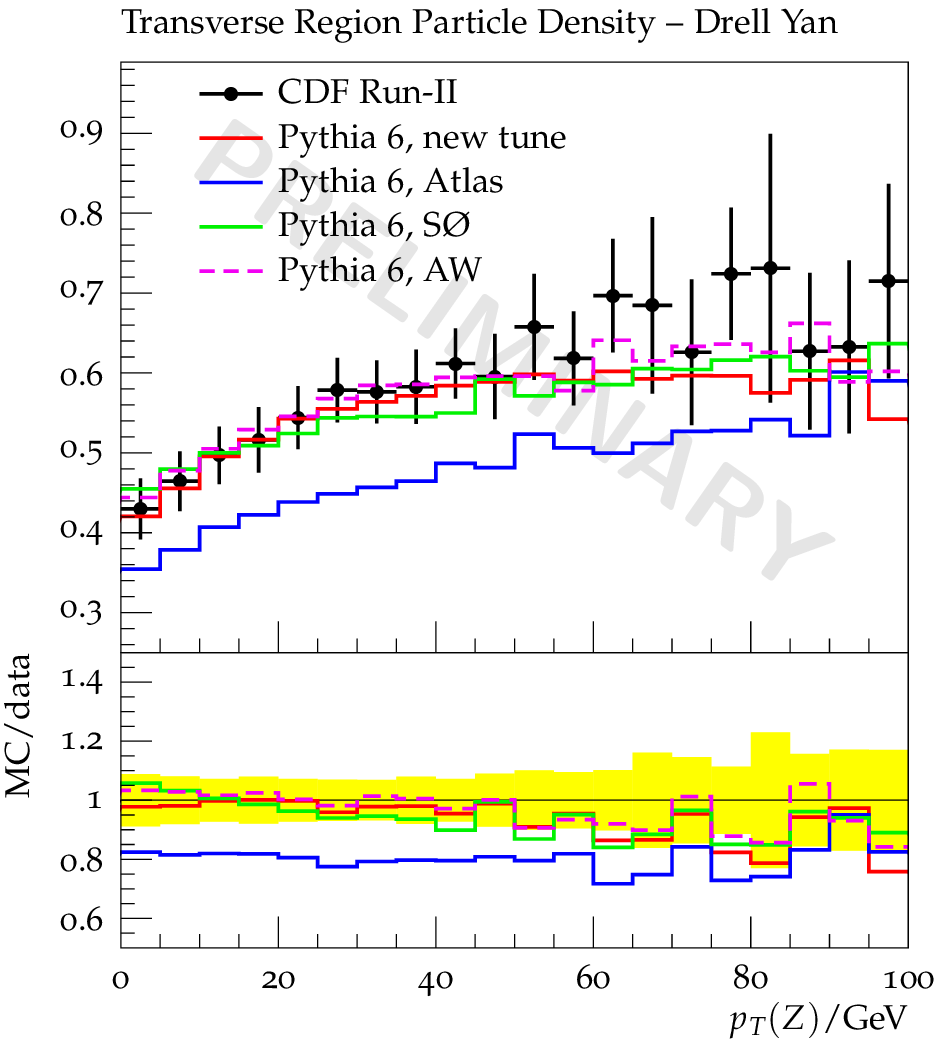}}%
  \qquad
  \subfloat[][]{\includegraphics[width=0.45\textwidth]{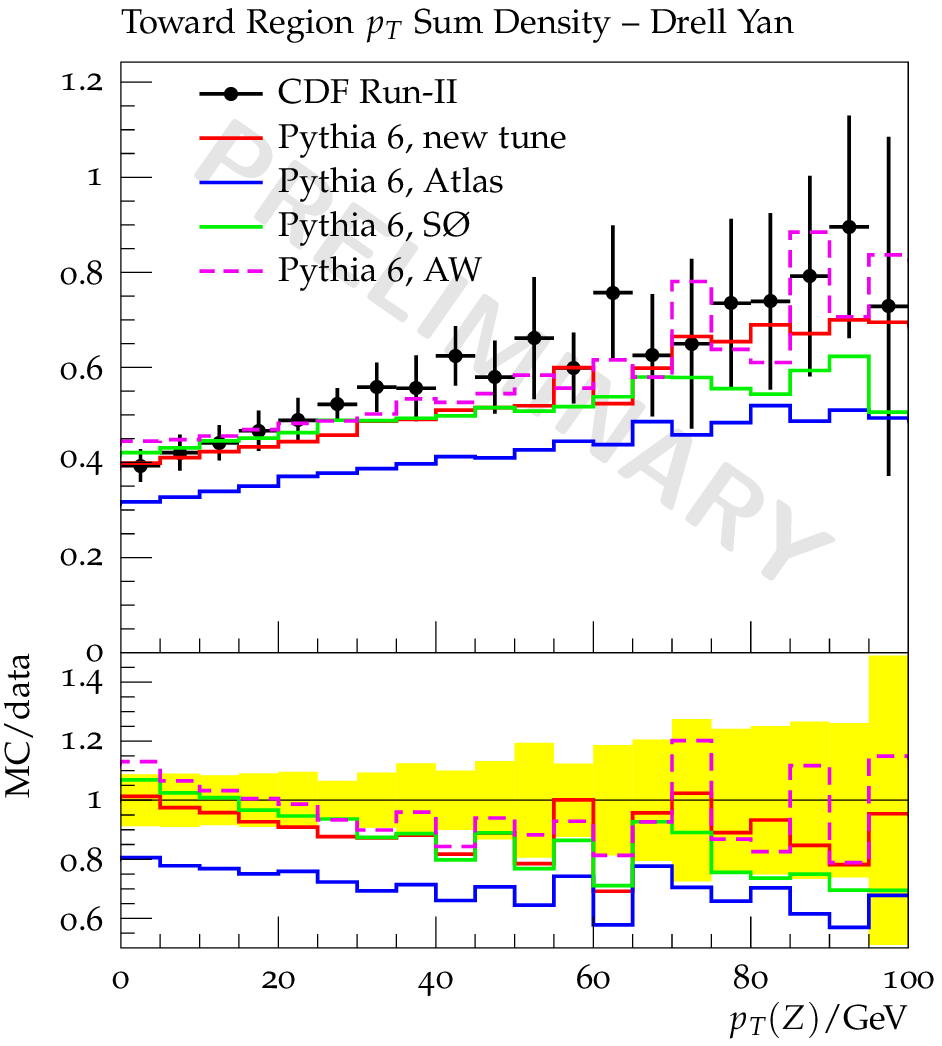}}%
  \qquad
  \subfloat[][]{\includegraphics[width=0.45\textwidth]{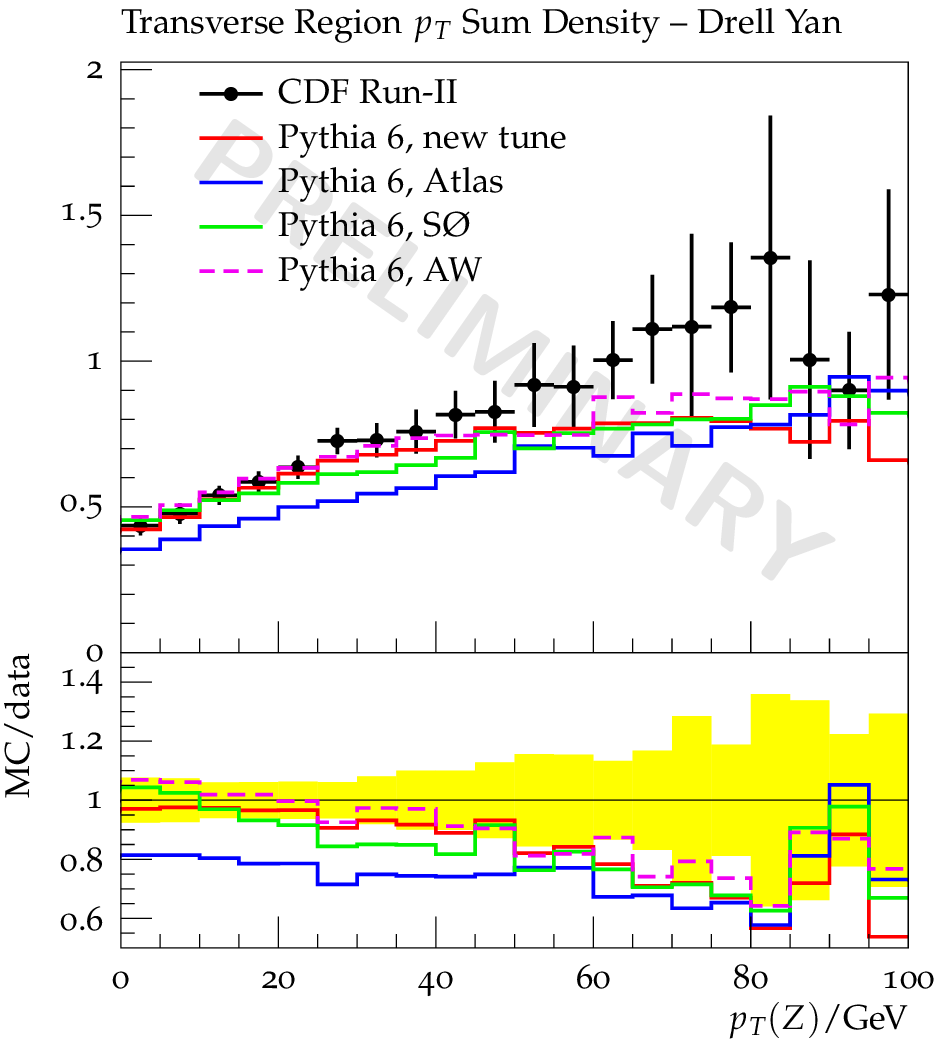}}%
  \caption{Pythia 6 ($Q^2$ shower, ``old'' MPI) comparative performance on Tevatron
    minimum bias and underlying event data at 1800 and \unit{1960}{\GeV}.}%
  \label{fig:pyppbarobs}%
\end{figure}


\clearpage

\subsection{Comparisons of various generator descriptions of tune observables}

For interest, we also show comparisons of the default tunes of several
generators on the same set of analysis distributions. This is of interest in
that it highlights the degree of variation possible between very similar models
and tunes: it is clear that for LHC purposes the tuning of many generators other
than Pythia~6 has a lot of room for improvement. All generators shown in the
plots of Figure~\ref{fig:miscobs}, with the exception of the Pythia~6 ``AW''
tune, are incapable of describing minimum bias data, and hence are cut off below
a leading jet $p_\perp$ of \unit{30}{\GeV} for the QCD distributions. 

Fortran Herwig\,\cite{Corcella:2002jc} (plus the
Jimmy\,\cite{Butterworth:1996zw} UE model) has been shown both with the Jimmy
default and the Atlas tune, but obviously suffers from not having been tuned as
extensively by LEP and Tevatron experiments as Pythia~6. This should be of
concern to Atlas, who are using this generator for SUSY simulation due to its
incorporation of spin correlations in SUSY decays.

Herwig++\,\cite{Bahr:2008tf} has been brute-force tuned to the \CDF 2001 jets UE
analysis, and hence its fit is rather good for most
distributions. Interestingly, the \CDF 2008 Drell-Yan UE study, which was not
available for the brute-force tuning, shows a deficiency in the UE model tune:
this data appears to break a degeneracy between several ways of obtaining good
fits to initial state data, and will be addressed by a \Professor tune of
Herwig++. Similarly Sherpa\,\cite{Gleisberg:2008ta}, which has only been tuned
``by eye'' so far, will undergo a \Professor tune in the near future.

\begin{figure}[b]%
  \centering
  \subfloat[][]{\includegraphics[width=0.45\textwidth]{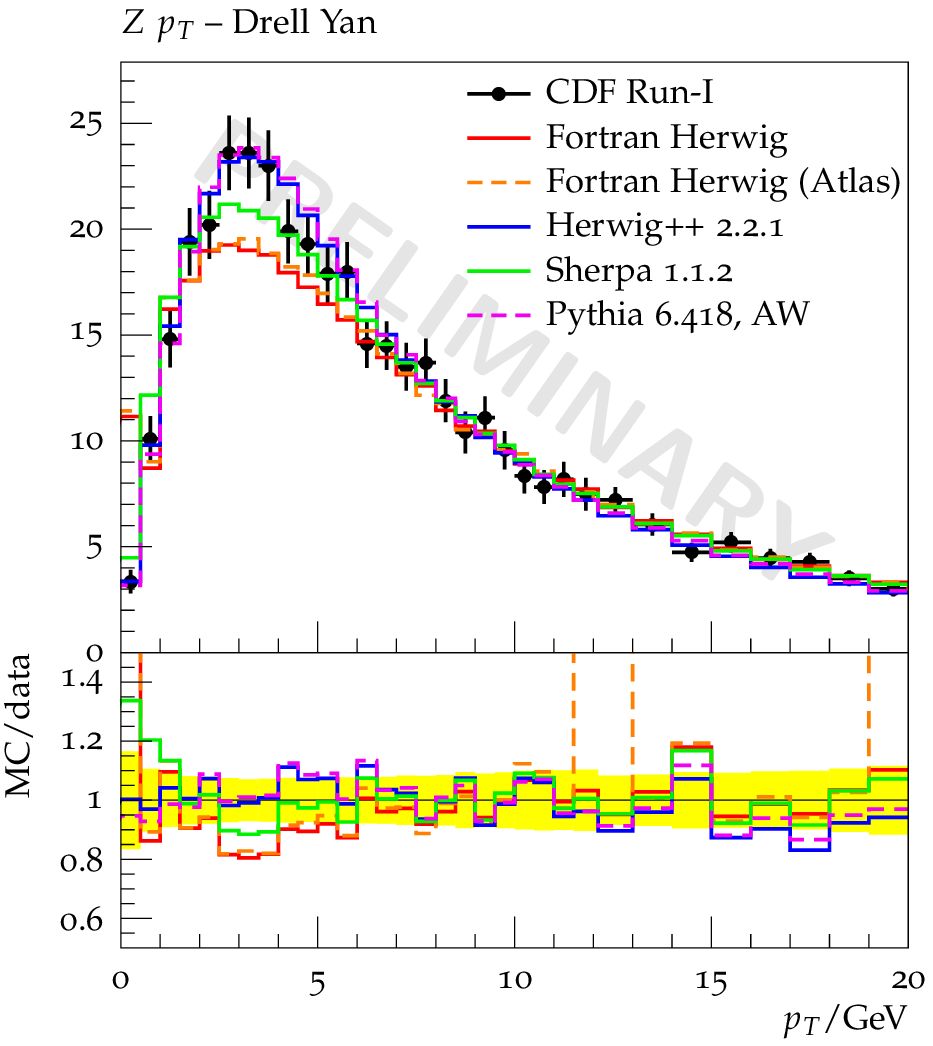}}%
  \qquad
  \subfloat[][]{\includegraphics[width=0.45\textwidth]{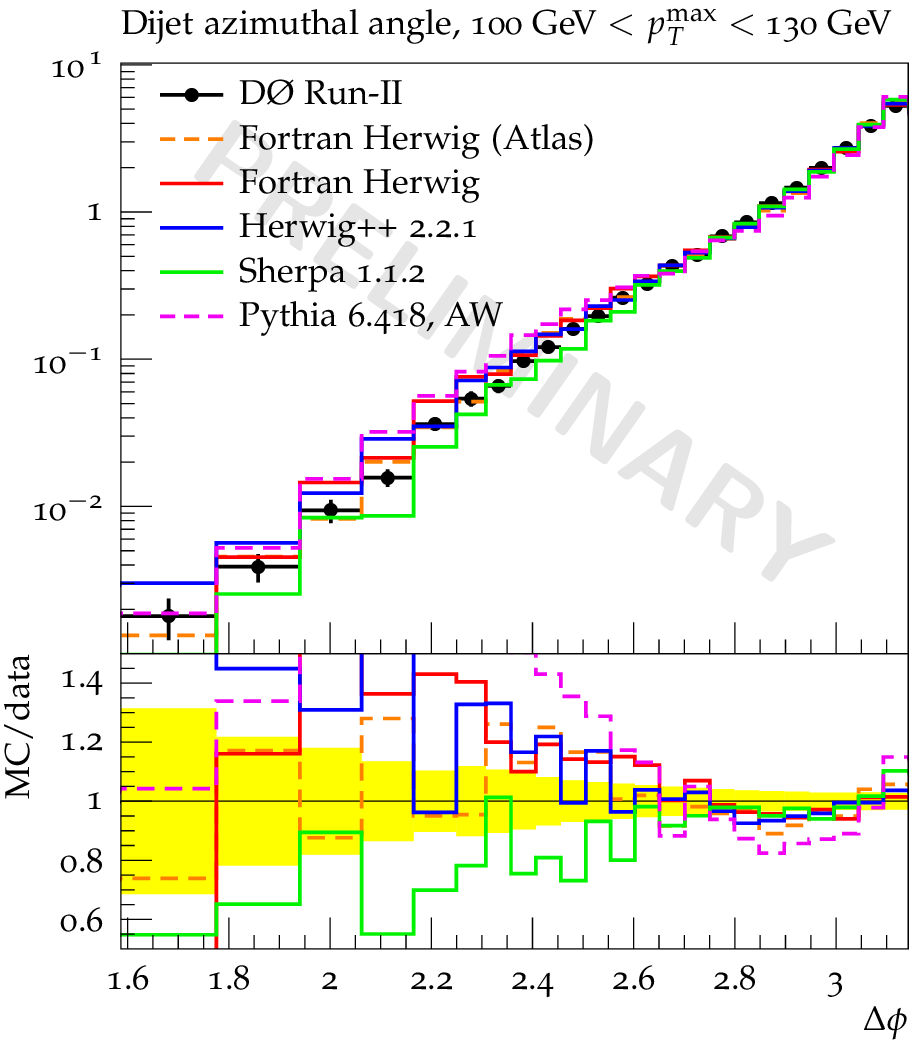}}%
  \caption{Comparisons of several LO MC event generators on Tevatron minimum
    bias and underlying event analyses at 1800 and \unit{1960}{\GeV}.}%
  \label{fig:miscobs}%
\end{figure}

\begin{figure}[p]%
  \ContinuedFloat
  \centering
  \subfloat[][]{\includegraphics[width=0.45\textwidth]{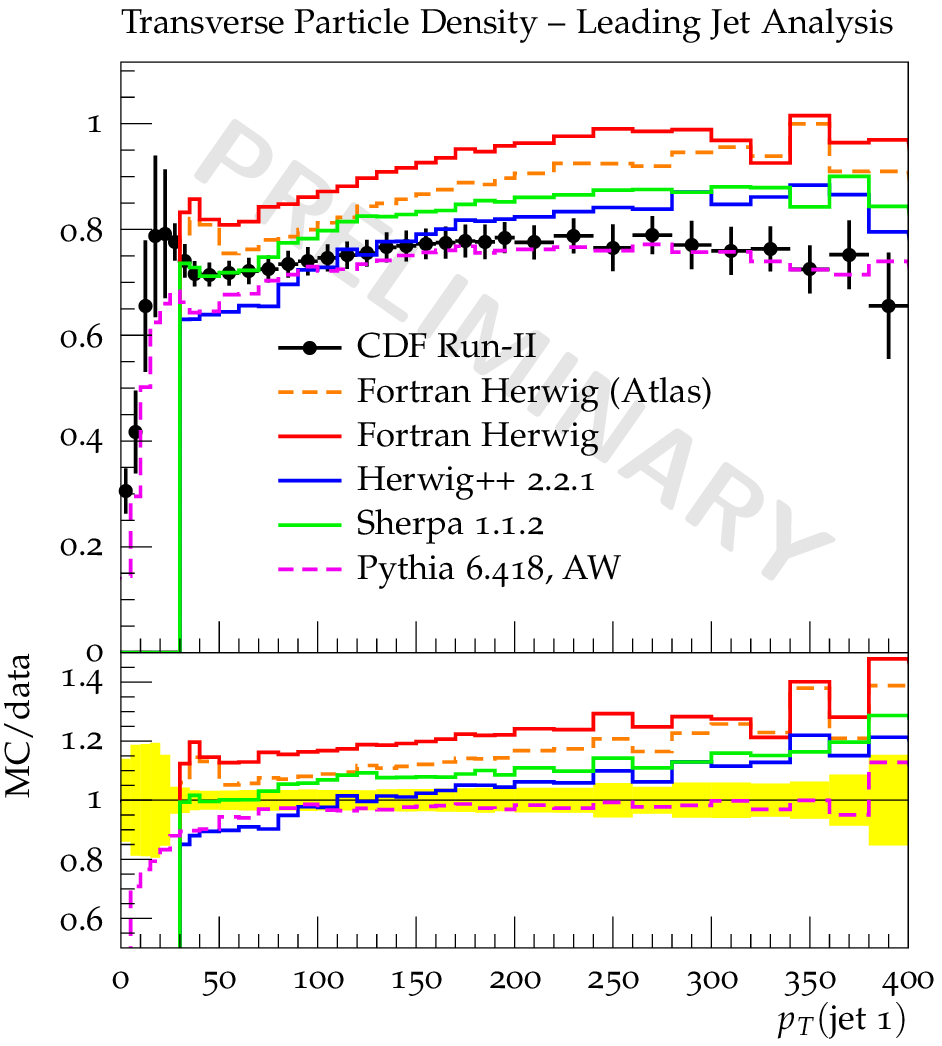}}%
  \qquad
  \subfloat[][]{\includegraphics[width=0.45\textwidth]{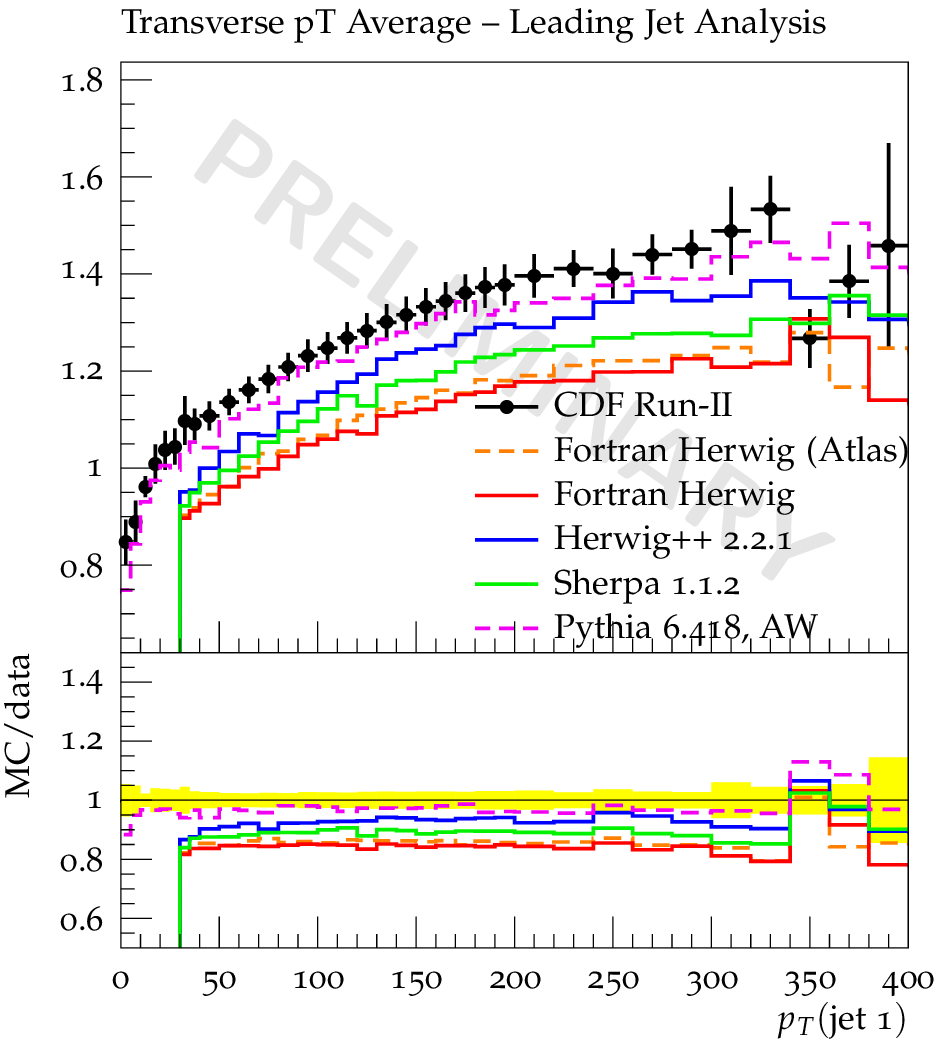}}%
  \qquad
  \subfloat[][]{\includegraphics[width=0.45\textwidth]{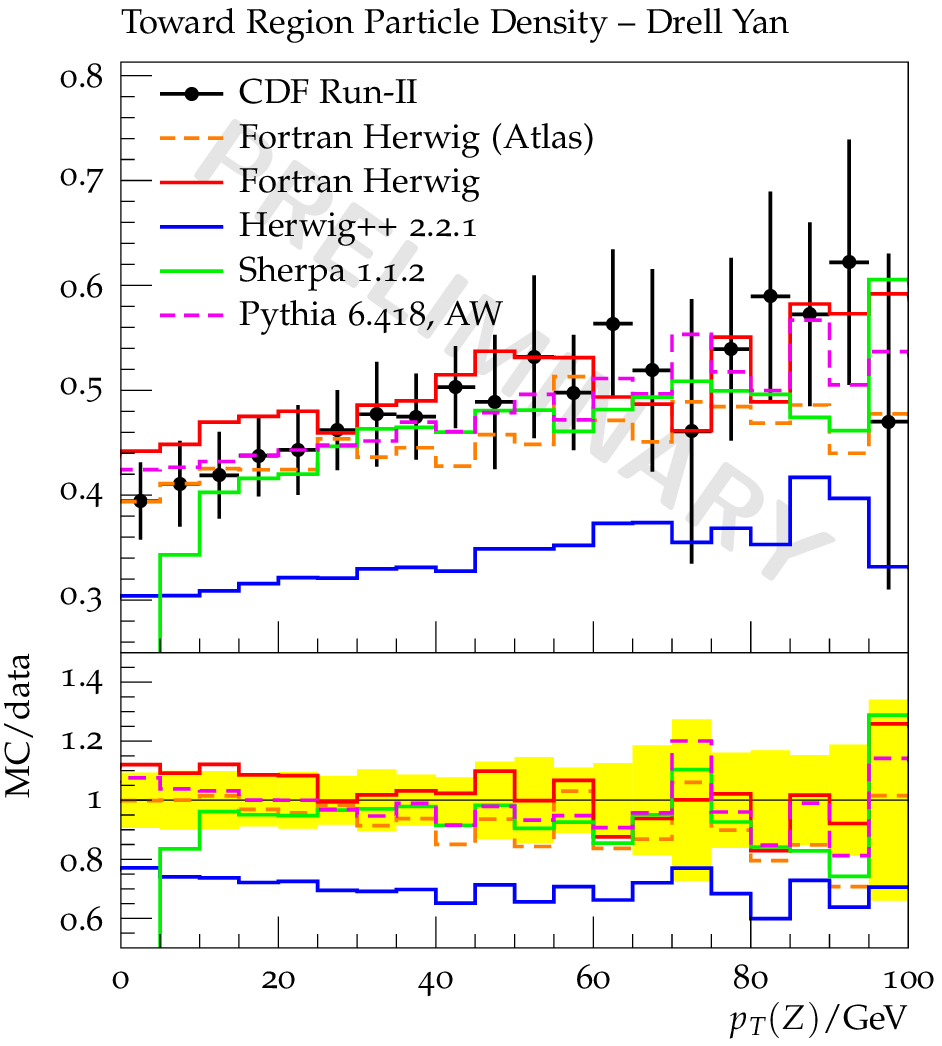}}%
  \qquad
  \subfloat[][]{\includegraphics[width=0.45\textwidth]{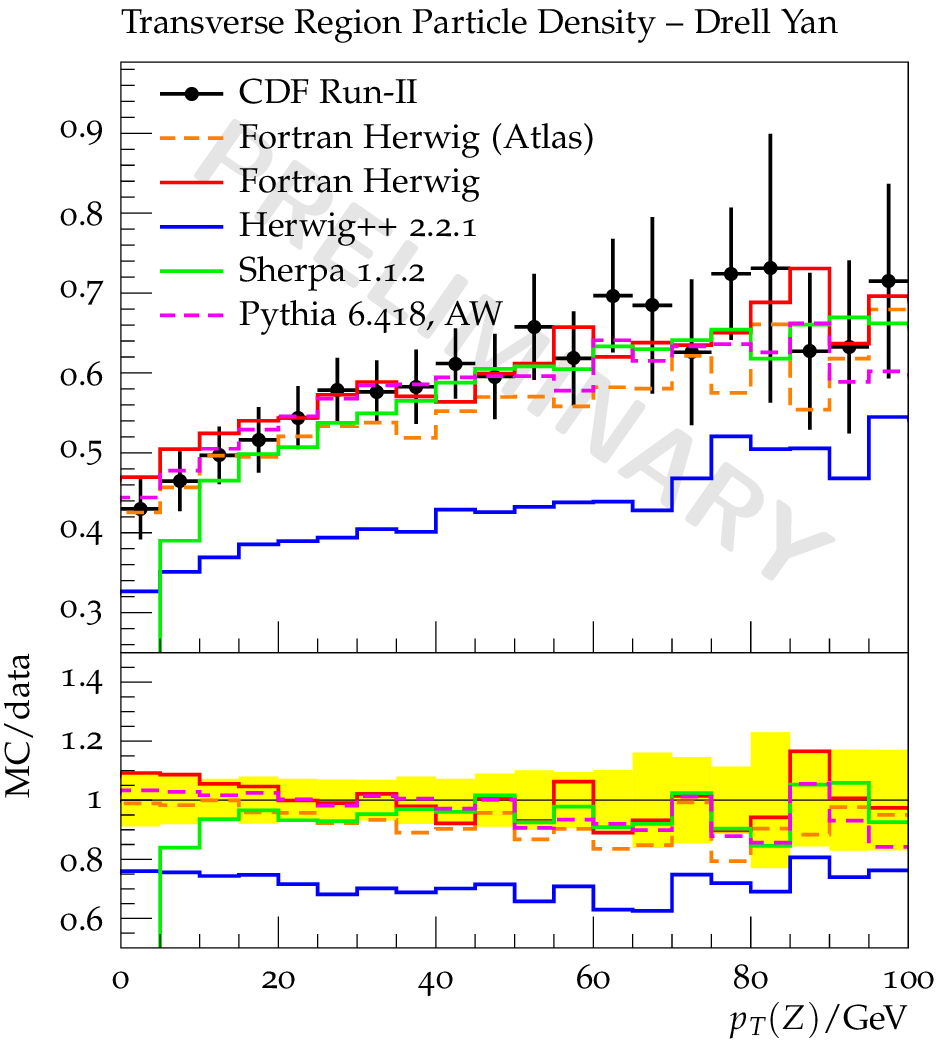}}%
  \caption{Comparisons of several LO MC event generators on Tevatron minimum
    bias and underlying event analyses at 1800 and \unit{1960}{\GeV}.}%
  \label{fig:miscobs}%
\end{figure}

\begin{figure}[p]%
  \ContinuedFloat
  \centering
  \subfloat[][]{\includegraphics[width=0.45\textwidth]{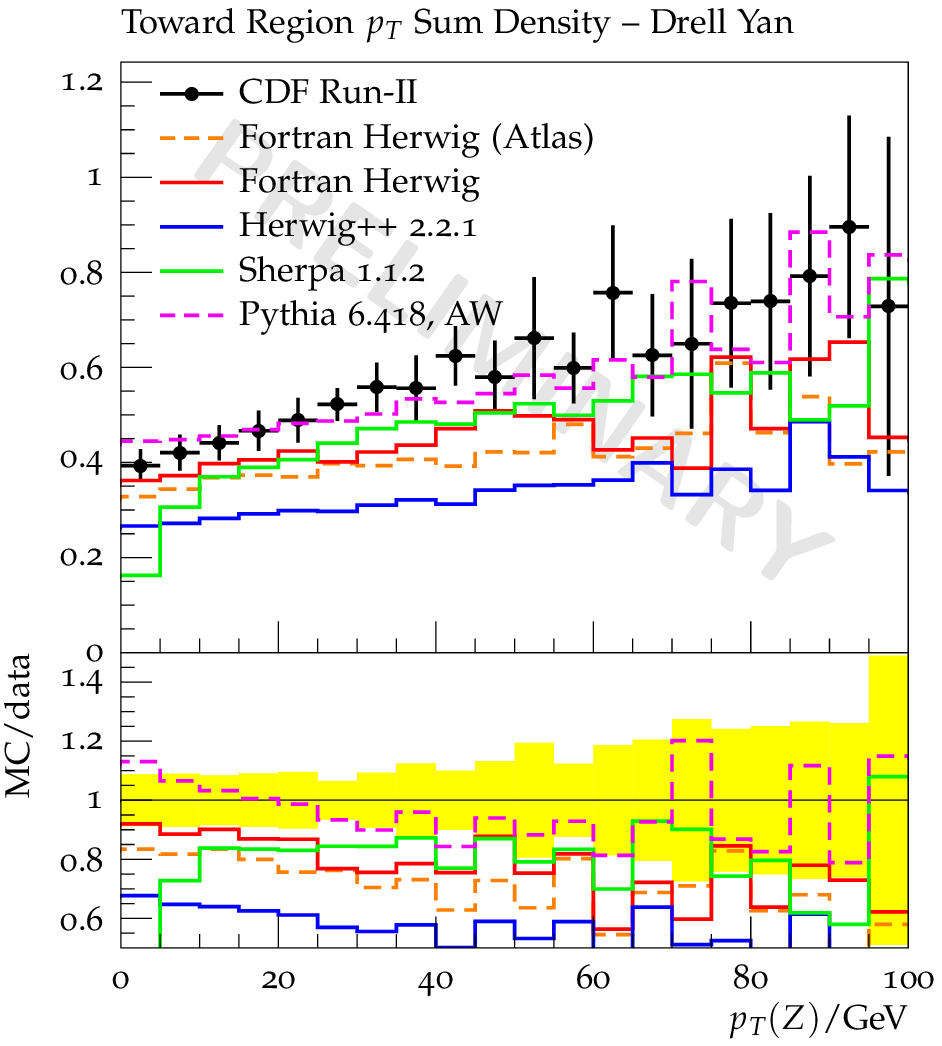}}%
  \qquad
  \subfloat[][]{\includegraphics[width=0.45\textwidth]{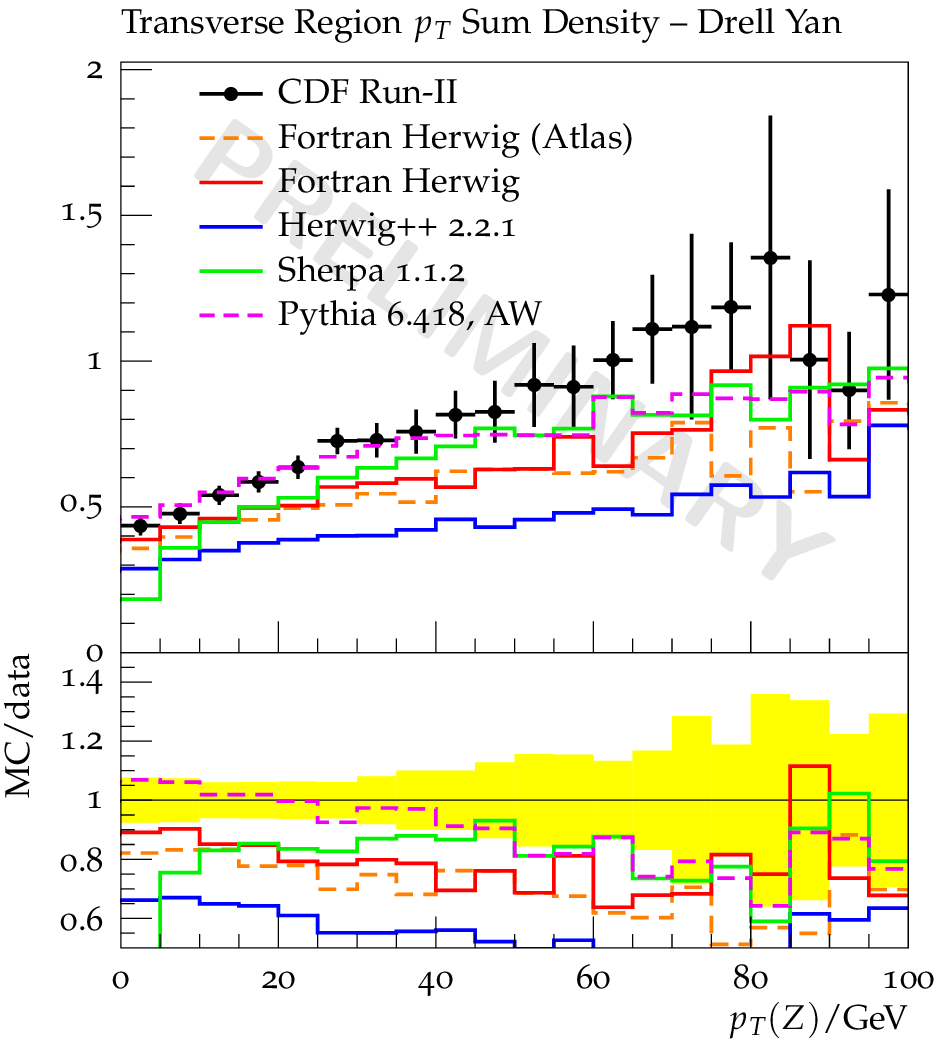}}%
  \qquad
  \subfloat[][]{\includegraphics[width=0.45\textwidth]{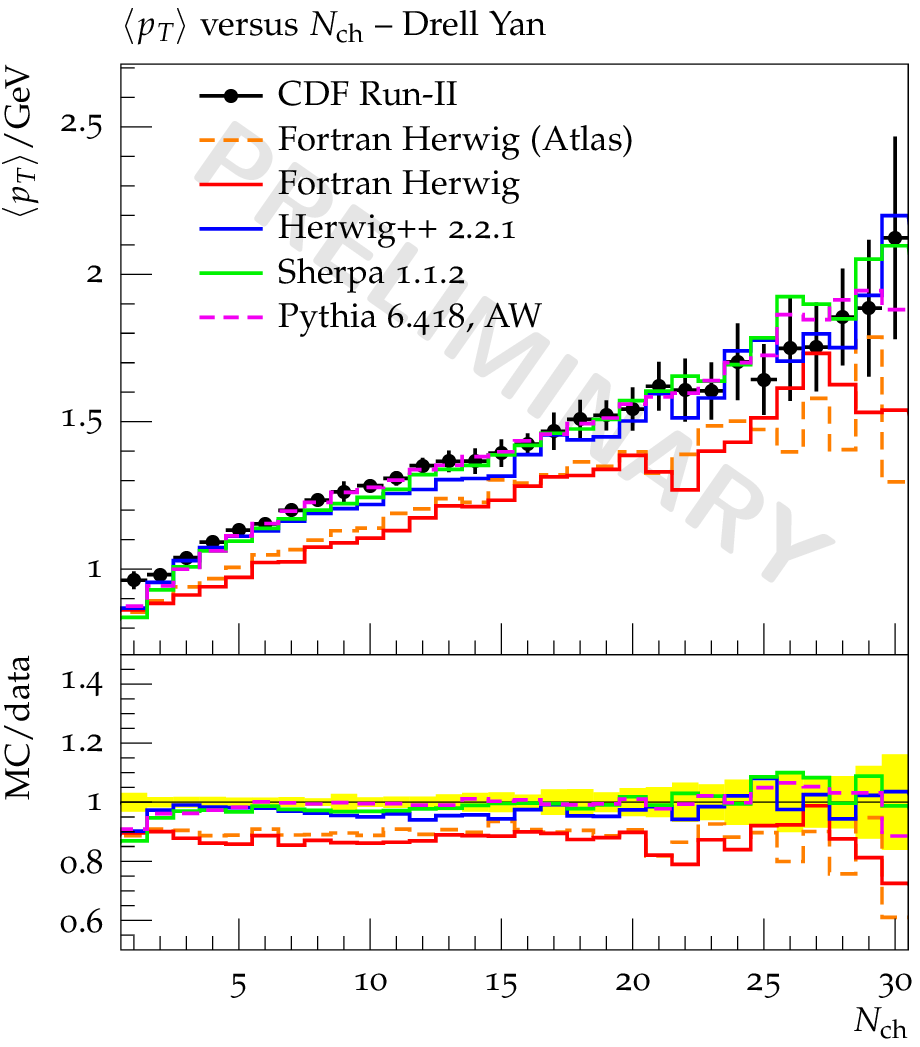}}%
  \qquad
  \subfloat[][]{\includegraphics[width=0.45\textwidth]{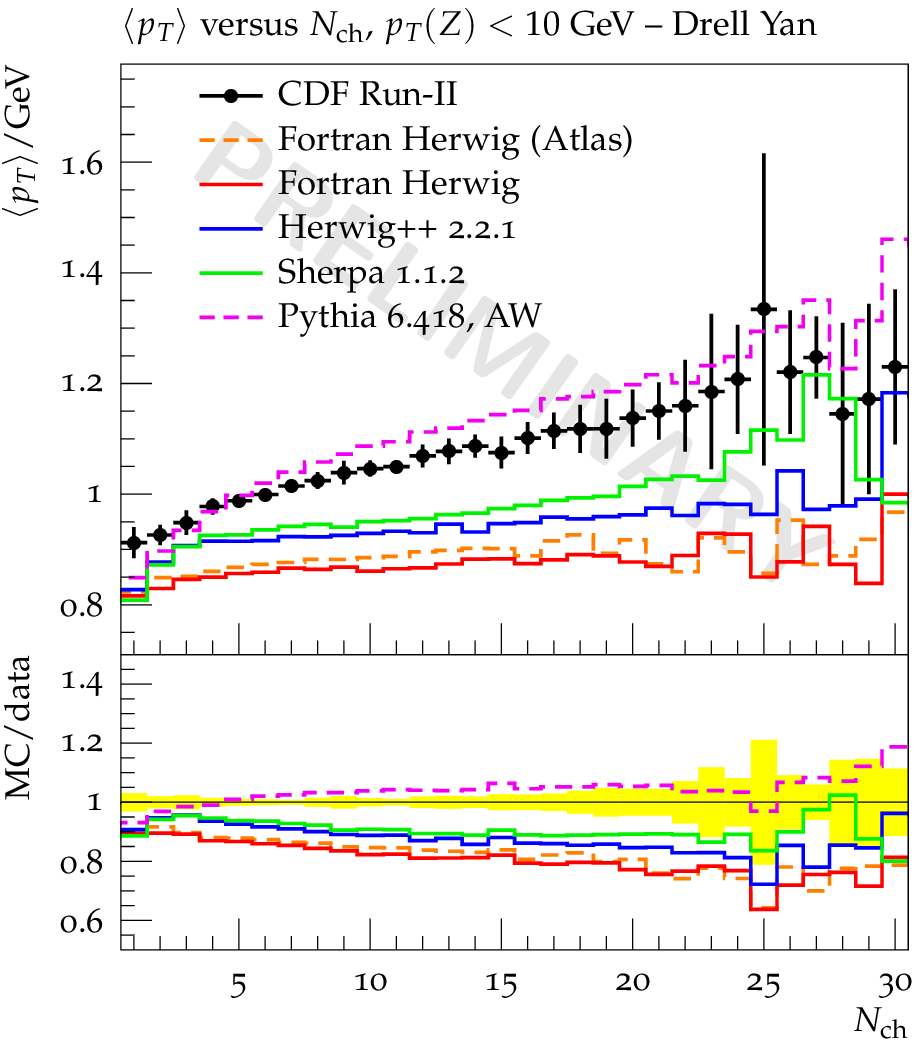}}%
  \caption{Comparisons of several LO MC event generators on Tevatron minimum
    bias and underlying event analyses at 1800 and \unit{1960}{\GeV}.}%
  \label{fig:miscobs}%
\end{figure}


\clearpage

\section{Conclusions}

We have summarised the work on Monte Carlo event generator validation and tuning
that has been done in the past year by the CEDAR collaboration, now integrated
in the MCnet research network. The most visible aspects of this development
work are the \Rivet validation analysis system and the \Professor tuning
system. The latter is an extension of earlier systematic tuning work, and uses
fitted bin-by-bin parameterisations of data response functions to predict the
overall goodness of fit to experimental data, from which point a numerical
optimisation is tractable.

Using the \Professor system to tune data produced by \Rivet from events
generated by the Pythia~6 generator code (in the $Q^2$-ordered parton shower
mode), we have obtained a tune of Pythia~6 which combines a good description of
data from LEP to the Tevatron. This tune has been extended to the Pythia~6
$p_\perp$-ordered parton shower/interleaved MPI model and will be documented in
a forthcoming publication. Extensions of the \Rivet library to data from the
RHIC, SPPS and ISR colliders will allow for more extensive tunes, in particular
constraining the evolution of the total $\Pproton\Pproton$/$\Pproton\APproton$
QCD cross-section, an important feature of minimum bias and underlying event
modelling for the LHC experiments.

The MC generator and SM groups on Atlas and CMS are currently beginning to use
\Rivet for MC validation and \Professor tunes will be provided for the main
generators --- Pythia~6, Herwig~6/Jimmy, Sherpa, Herwig++ and Pythia~8 --- to be
used as base configurations by both collaborations. This work is just beginning,
and the details depend on the choice of PDFs for experiment LO generator
production.

The phenomenological nature of low-$p_\perp$ QCD modelling means that even the
best fits to UE energy extrapolation can be disrupted by new data at LHC
energies: the effect of Drell-Yan UE data on the otherwise good fits of the
Herwig++ and Atlas Pythia~6 tunes stands as testament to the ability of new data
to surprise. Accordingly, we intend for \Professor to be available for private
use within the LHC collaborations, to allow rapid response of generator tunes to
first QCD data. This updating is necessary for good understanding of LHC QCD
backgrounds to BSM physics, and is hence in every experimentalist's
interest. The next year will see many physics surprises: our hope is that with
systematic tools to evaluate and improve their behaviours, the MCnet Monte Carlo
event generators will prove up to the task.

\section*{Acknowledgements}

We would like to thank Frank Krauss for convening the \Professor collaboration
and for useful discussions; a tune of Sherpa is the least we can do in return!
This work was supported in part by the MCnet European Union Marie Curie Research
Training Network, which provided funding for collaboration meetings and
attendance at research workshops such as ACAT08. AB has been principally
supported by a Special Project Grant from the UK Science \& Technology Funding
Council. HH acknowledges a MCnet postdoctoral fellowship.